\documentclass[prd,singlecolumn,amsmath,amssymb,nobibnotes,nofootinbib,superscriptaddress,preprintnumbers]{revtex4}
\usepackage{graphicx}

\newcommand{\nn}{\nonumber}
\newcommand{\be}{\begin{equation}}
\newcommand{\ee}{\end{equation}}
\newcommand{\bea}{\begin{eqnarray}}
\newcommand{\eea}{\end{eqnarray}}

\def\a{\alpha}
\def\b{\beta}

\def\d{\delta}

\def\s{\sigma}

\def\e{\epsilon}

\def\th{\theta}

\def\s{\sigma}

\def\t{\tau}

\def\ie{{\it i.e. }}

\begin{document}

\title{Cycles in the Multiverse}

\author{Matthew C. Johnson}
\affiliation{Perimeter Institute for Theoretical Physics, 31 Caroline St N, Waterloo, Ontario N2L 2Y5, Canada}
\author{Jean-Luc Lehners}
\affiliation{Max-Planck-Institute for Gravitational Physics (Albert-Einstein-Institute), D-14476 Potsdam/Golm, Germany}

\begin{abstract}
Eternal inflation is a seemingly generic consequence of theories that give rise to accelerated expansion of the universe and possess multiple vacuum states. Making predictions in an eternally inflating universe is notoriously difficult because one must compare infinite quantities, and a wide variety of regulating procedures yield radically different results. This is the measure problem of eternal inflation. In this paper, we analyze models of eternal inflation which allow for the possibility of cyclic bubble universes: in each bubble, standard cosmological evolution is re-played over and over again. Eternal inflation can generically arise in cyclic models that include a dark energy dominated phase. In such models, several problematic consequences of standard regulating procedures, such as the youngness and Boltzmann Brain problems, are substantially alleviated. We discuss the implications for making predictions in cyclic models, as well as some general implications for understanding the measure problem in eternal inflation.
\end{abstract}

\maketitle

\section{Introduction}

In the standard cosmological model, accelerated expansion is invoked at least twice to explain our observations of the cosmos: dark energy in the late universe, and cosmic inflation in the early universe. This theory allows multiple phases, characterized by vastly different rates of accelerated expansion; inflation gave way to a dark energy dominated universe with an expansion rate some $50$ orders of magnitude smaller. In fact, the possibilities may be much more varied, since modern theories of particle physics relying on spontaneous symmetry breaking generically do not predict a unique vacuum state, let alone a unique cosmological history.

Many such theories that allow multiple phases (or vacua) and accelerated expansion give rise to a phenomenon known as eternal inflation. In the simplest example of eternal inflation, different vacua are accessed by a first order phase transition. In this picture, localized pockets containing a new phase are formed from a ``parent" vacuum undergoing accelerated expansion. These pockets subsequently expand, but if they are formed at a rate slower than the rate of expansion, the phase transition cannot complete, and inflation becomes eternal. The original inflating vacuum continues to grow and spawn regions of all the possible phases (for a review of eternal inflation, see e.g.~\cite{Aguirre:2007gy}). In this way, eternal inflation makes all possibilities a reality.

There are many conceptual problems that must be confronted in order to make sense of eternal inflation. Since the many possibilities allowed by an underlying theory are {\em actually} realized somewhere in spacetime, predictions for what we might observe become intrinsically statistical. In a finite universe with a finite number of possibilities, this would not be a problem. However, in an eternally inflating universe each possibility is realized an infinite number of times, and so a frequentist's definition of probabilities becomes ill-defined without some prescription to regulate the infinities. The most straightforward  procedure is to use a geometrical cutoff: count the frequency of some occurrence vs others at a finite time (say, by counting the volume occupied by different phases), and then take the time at which the ratio is evaluated to infinity. However, this procedure is not unique, and choosing different timeslicings yields different sets of probabilities. This is an example of the ``measure problem" of eternal inflation.

To the extent that they are understood, the predictions made by various measures (cutoff procedures) range from being absurdly in conflict to fairly consistent with data (at least for a few variables such as the cosmological constant). For example, the generic prediction of measures that weight by physical volume is that we should be as ``young" as possible (i.e. everything else constant, we should exist at the earliest possible stage in the evolution of the universe consistent with our existence). This is known as the Youngness Problem. If there is a cosmological constant in each pocket (as there is evidence for in our universe), then because of the thermodynamic properties of de Sitter space, many measures predict that we should be ``freak observers'' (i.e. that we didn't arise from our observed cosmological evolution) formed by a thermal fluctuation out of the vacuum at some very late time. This is known as the Boltzmann Brain Problem. We will discuss both of these issues at length below.

One strategy for making progress on the measure problem is to rule out those measures which make predictions that do not agree with data, or are not internally consistent (see e.g. the discussion in Ref.~\cite{Bousso:2008hz}). For example, one would rule out those measures with a Youngness or Boltzmann Brain Problem. Another approach is to look for guidance from fundamental theory, which perhaps would single out the correct prescription~\cite{Bousso:2010id,Vilenkin:2011yx,Garriga:2008ks,Sekino:2009kv,Freivogel:2006xu}. However, this is a complex problem which is not yet resolved.

An alternative to the inflationary epoch in the standard cosmological model is the ekpyrotic/cyclic universe scenario~\cite{Khoury:2001wf,Steinhardt:2001st,Steinhardt:2002ih,Khoury:2001bz,Steinhardt:2004gk}, for a review see~\cite{Lehners:2008vx}. This scenario purports to address both the homogeneity and isotropy of our observed universe as well as seeding structure with appropriate density fluctuations. This is accomplished by invoking an ``ekpyrotic'' phase in the early universe, in which there is slow contraction accompanied by a rapid change in the energy density. This contraction gives way to expansion after a transition from big-crunch to big-bang, after which the standard big-bang cosmology follows. Far in the future, the universe will contract again, and the cycle repeats itself. This picture is motivated by Heterotic M-theory, as we describe in greater detail below.

\begin{figure}
\begin{center}
\includegraphics[width=0.75\textwidth]{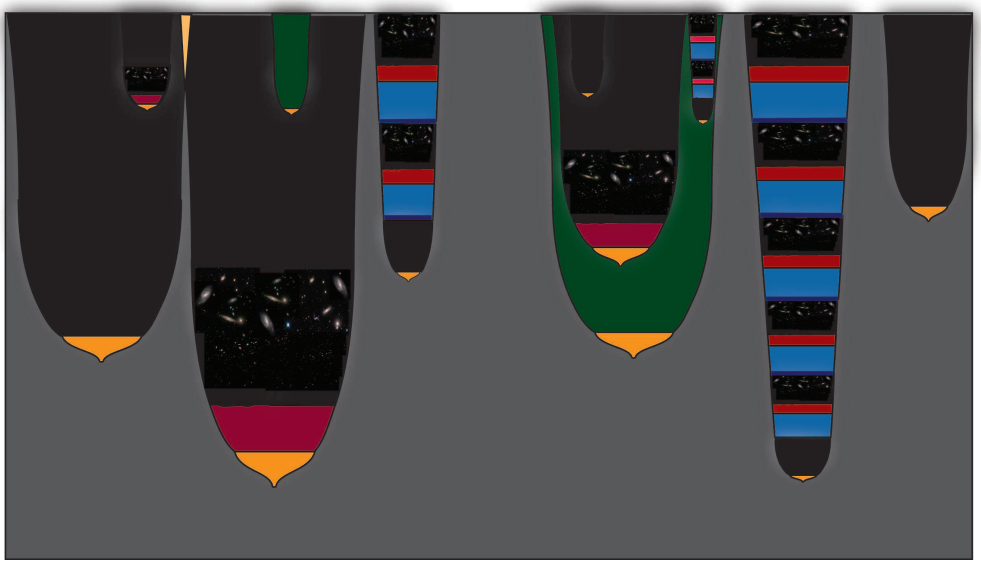}
\caption{The Cyclic/Inflationary Multiverse.
\label{Fig-Multiverse} }
\end{center}
\end{figure}

There are at least two reasons why one must address the question of measure in the cyclic universe. The first arises when there can be differences between cycles, as there are in the so-called Phoenix Universe~\cite{Lehners:2008qe,Lehners:2010ug,Lehners:2011ig}. In this setup, the amplitude of density fluctuations can vary from place to place in each cycle. However, only those parts of the universe that are sufficiently empty and flat make it from the contracting to the expanding phase, leading to a selection factor for the amplitude of density perturbations. Different possibilities are sampled in different spatiotemporal regions, requiring one to define a statistical measure for predictions (and in an infinite universe, some cutoff procedure). Second, the cyclic universe scenario invokes accelerated expansion to smooth the universe prior to ekpyrotic contraction. As argued above, once accelerated expansion is added to any cosmological model, the existence of eternal inflation becomes a fairly generic possibility. This is even more pronounced given that the cyclic universe is motivated by M-theory, where the many ways to compactify extra dimensions give rise to many possible four-dimensional theories. The picture of the multiverse might then contain both cyclic and inflationary cosmologies, as shown in Fig.~\ref{Fig-Multiverse}.

Can the cyclic universe arise in local pockets spawned from an eternally inflating universe? What are the implications for making predictions in such models? In this paper, we address these questions, finding that using existing measures the youngness and Boltzmann brain problems need not (under a set of conditions we outline in detail) arise in a theory where many cyclic universes are seeded by eternal inflation. In addition, we construct a model where both cyclic and standard inflationary bubbles coexist, and find under which circumstances these problems reappear. One should keep in mind our largest assumptions: 1) that there are multiple phases (including cyclic phases) which can be seeded by eternal inflation and 2) that the full theory allows a predictive transition from a big crunch to a big bang following the ekpyrotic phase. Both assumptions are plausible, however unproven; we offer no new critiques of either the eternal inflation or cyclic scenarios. Rather, we wish to illustrate how the inclusion of cyclic universes can lead to drastically different predictions for where we might find ourselves in the multiverse. Our results also highlight how different assumptions about the physics underlying the multiverse can render previously discarded measure prescriptions consistent with our observed universe.

The plan of the paper is as follows. In Sec.~\ref{Section-Setup}, we describe how eternal inflation might arise within the cyclic universe scenario. In Sec.~\ref{Section-RateEqs}, we outline how predictions are extracted, and how the eternally inflating cyclic universe can avoid the youngness and Boltzmann brain problems. We add inflationary bubbles to the setup in Sec.~\ref{sec:inflationarybubbles} and show under which circumstances these problems reappear, and then conclude in Sec.~\ref{sec:conclusions}. Finally, we include a discussion of tunneling to the cyclic universe in Appendix~\ref{sec:twofieldbubbles} and a calculation for worldline based measures in Appendix~\ref{sec:Boussorates}.

\section{The Cyclic Universe and Eternal Inflation} \label{Section-Setup}

\begin{figure}
\begin{center}
\includegraphics[width=0.75\textwidth]{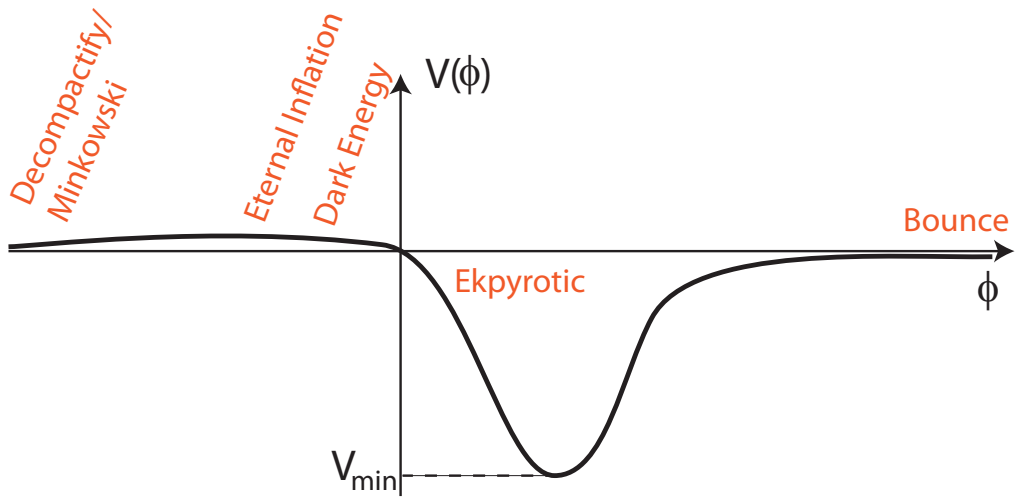}
\caption{A potential that gives rise to slow-roll eternal inflation and cyclic cosmologies. The inflating phase near the local maximum can exit either to a decompactified phase (left) or a cyclic pocket universe (right).
\label{Fig-CyclicPlusSREI} }
\end{center}
\end{figure}

The cyclic universe is inspired by the braneworld picture stemming from heterotic M-theory \cite{Horava:1995qa,Lukas:1998yy,Lukas:1998tt}. According to this picture, we live on a $(3+1)$-dimensional brane, and are separated by a fifth dimension from a second brane. Both of these branes are boundaries of the interior ``bulk'' spacetime. This model can also be embedded in eleven-dimensional supergravity, in which case there is an additional six-dimensional internal Calabi-Yau manifold at each point in the five-dimensional braneworld spacetime. In Ref.~\cite{Khoury:2001wf}, Khoury {\it et al.} speculated that there could be forces between the branes which would cause the boundary branes to slowly attract each other (these forces can be modeled by a potential for the inter-brane distance modulus $\phi$, see Fig~ \ref{Fig-CyclicPlusSREI}). From a four-dimensional effective point of view, such an {\it ekpyrotic} phase corresponds to a slow contraction of the universe. During this phase, quantum fluctuations can be amplified to classical perturbations, similarly to what happens in inflation. When the two branes eventually collide, matter and radiation are produced with the perturbations imprinted on them. This corresponds to the big bang, with the universe bouncing from a contracting to an expanding phase. The branes quickly separate again, but due to Hubble friction they come to a halt on the plateau marked ``dark energy'' in Fig. \ref{Fig-CyclicPlusSREI}. At this point, the ordinary phases of radiation and matter domination occur, until eventually the scalar field potential energy becomes the dominant component in the universe and a phase of dark energy ensues. This phase lasts until the scalar field rolls down to negative values of the potential, at which time the universe reverts from expansion to contraction, a new ekpyrotic phase takes place and a new cycle is underway \cite{Steinhardt:2001st}.

The region of the potential at large separation of the branes has not been much considered so far. In the standard setup, the field returns to the location marked ``dark energy" on the potential, but no farther. However, there is nothing preventing the branes from starting at large separation on the first cycle, or from rare quantum fluctuations pushing them far apart on subsequent cycles. It seems reasonable to assume that, if very large brane separations are allowed, the potential approaches zero asymptotically due to decreasing inter-brane forces (with our conventions in the figure, this corresponds to $V(\phi) \rightarrow 0$ as $\phi \rightarrow -\infty$; in this limit both the inter-brane dimension and the Calabi-Yau manifold become large). In this case, there is a local maximum of the potential at positive energy, and slow-roll eternal inflation can take place. Some regions of the universe will decompactify, while others will undergo ekpyrosis and repeated cycling.

\begin{figure}
\begin{center}
\includegraphics[width=0.75\textwidth]{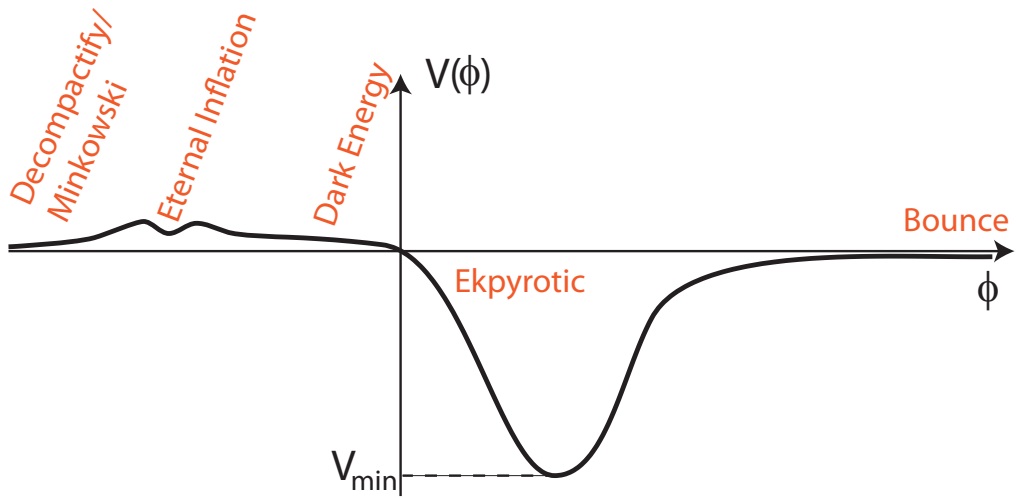}
\caption{The cyclic universe incorporating false vacuum eternal inflation.
\label{Fig-CyclicPlusFVEI} }
\end{center}
\end{figure}

It is also possible that the brane potential has a local minimum above the dark energy plateau as shown in Fig. \ref{Fig-CyclicPlusFVEI}. In that case, false vacuum eternal inflation occurs when the field resides in this local minimum. From the false vacuum, tunneling to either side of the potential barrier is mediated by the Coleman-de Luccia (CDL) instanton~\cite{Coleman:1980aw}. Tunneling corresponds in real space to the nucleation of an expanding bubble  containing a new phase embedded in the false vacuum. Inside each bubble is an infinite open Friedman Robertson Walker (FRW) universe. If the tunneling occurs to the left, inside the new bubble universe the scalar will roll off to $-\infty,$ corresponding to the spacetime decompactifying. The second possibility is tunneling to the right, which corresponds to the nucleation of a {\it cyclic bubble}. After an initial period of curvature domination, a dark energy dominated phase ensues which dilutes the curvature and any initial inhomogeneities. The dark energy phase is followed by the ekpyrotic contraction phase during which perturbations are generated and classical anisotropies are further suppressed \cite{Erickson:2003zm}. As described above, after the ekpyrotic phase the bubble interior undergoes a crunch/bang transition, accompanied by the creation of radiation and matter \cite{Turok:2004gb}. During the subsequent phases of radiation and matter domination stars, galaxies and ordinary observers are formed, the dark energy subsequently takes over for many billions of years and then a new cycle starts.

The original cyclic universe \cite{Steinhardt:2001st} assumed that there is a single scalar field that is responsible both for the background dynamics and whose quantum fluctuations are amplified into scale-invariant density perturbations. In that case, the scale factor grows by $60+N_{de}$ e-folds over the course of one cycle, the $60$ e-folds stemming from the growth during the radiation and matter phases, and the $N_{de}$ e-folds from the dark energy phase. Note that the total contraction during the ekpyrotic phase is in fact negligible. Hence, in this case, it is clear that, even for a short dark energy phase, the universe grows hugely from cycle to cycle. However, it remains unclear whether this model actually produces a scale-invariant spectrum of perturbations, as this depends on the matching conditions at the bounce \cite{Creminelli:2004jg}. The perturbations that are present right before the bounce have a very blue spectrum \cite{Khoury:2001zk,Lyth:2001pf,Tsujikawa:2002qc}, and would not be consistent with observation. Under certain assumptions about the physics of the bounce \cite{Tolley:2003nx} (namely, that the evolution remains unitary, though becomes non-local), these blue perturbations can become scale-invariant, but it is important to remember that these assumptions remain unproven at present.

Two-field ekpyrotic models \cite{Lehners:2006pu,Lehners:2006ir} can produce scale-invariant curvature perturbations already prior to the bounce, via a curvaton-like mechanism where scale-invariant isocurvature/entropy perturbations are generated first, and converted into scale-invariant curvature perturbations just before the bounce \cite{Lehners:2007ac}. These models produce the required density perturbations more robustly, but at the expense of an instability: the two-field ekpyrotic potential is tachyonic transverse to the background trajectory in scalar field space. This instability means that only a portion of the spacetime experiences a full ekpyrotic phase, while the remaining regions become highly curved and undergo a fatal mixmaster crunch from which they can presumably not re-emerge \cite{Lehners:2008qe}. In the present paper, we are not interested in the details of this process -- rather, we will simply model this case by introducing a parameter $p$ describing the probability that a given spacetime region will make it through the bounce and into a new cycle. Thus, the condition that the two-field cyclic universe grows from cycle to cycle translates into the requirement that $p \, e^{180+3N_{de}}$ must be greater than one. Apart from this effect, the two-field cyclic universe can be treated identically to the single-field case. In particular, the second scalar field does not play a prominent role during the tunneling process, as we show in detail in Appendix A.

In addition to the specific scenarios outlined above, it should be remembered that the braneworld picture arises from the compactification of extra dimensions, which implies the possible existence of a vast landscape of possible four-dimensional vacua \cite{Susskind:2003kw}. If each of these vacua can be accessed from an eternally inflating parent vacuum, then it is certainly possible that bubbles containing an inflationary universe co-exist with bubbles containing a cyclic  cosmology, with each being realized in different regions. If both scenarios are shown to reproduce our observations, we would have to compare the two theories in a very real sense and ask: does the multiverse prefer one cosmology over another?

\section{Mapping the cyclic multiverse} \label{Section-RateEqs}

Let us be specific, and define what it means to make predictions in a multiverse (for a detailed discussion, see e.g.~\cite{Aguirre:2005cj}). A relevant quantity is the probability that a randomly chosen object $X$ (e.g. $X \equiv $ galaxy, baryons, etc.) experiences a set of parameters $\alpha$ (e.g. $\alpha \equiv $ cosmological constant, fine structure constant, etc.). It is convenient to split this quantity into two pieces:
\begin{equation}
\mathcal{P}_X (\vec{ \alpha}) \propto P_q (\vec{ \alpha}) n_{X,q} (\vec{ \alpha})
\end{equation}
where $P_q (\vec{ \alpha})$ is the ``prior'' probability that an object $q$ is associated with $\vec{ \alpha}$, and $n_{X,q} (\vec{ \alpha}) $ is the ``conditionalization:" the number of $X$ associated with $q$. For example, $q$ could be ``unit physical volume," in which case $P_q (\vec{ \alpha})$ is simply the fraction of the total volume having a set of parameters $\vec{ \alpha}$. The prior can therefore be thought of as a map of the multiverse, and will be the quantity we are primarily interested in for the remainder of this paper. We will refer to a choice of $q$ along with a prescription for calculating probabilities as a ``measure." The conditionalization is required to make connection with observation. That is, a particular set of parameters will be observed only if there are observers. Aside from informing our choice of prior, we do not consider the conditionalization further.

There are two broad choices of measure we consider in this paper: {\it global} and {\it local}. Global measures assign probabilities by comparing e.g. the relative proportion of volume taken up by various vacua~\cite{Garriga:2001ri,Linde:1993xx} or number of bubbles of each type~\cite{Garriga:2005av}. The idea is simply that the more of a given vacuum exists, the more likely it is to find oneself in that vacuum. Local measures follow the worldline of a single observer, and count histories~\cite{Aguirre:2006na} or the number of times a given vacuum is entered~\cite{Bousso:2006ev,Vanchurin:2006qp}. In some cases, the global and local measures are equivalent~\cite{Bousso:2009mw}. See e.g. Refs.~\cite{Vilenkin:2006xv,Aguirre:2006ak,Winitzki:2006rn,Freivogel:2011eg} for a general discussion of the various measure proposals.

\subsection{Rate equations}

The rate equations describe how the volume $f_i$ of vacuum $i$ (where the index $i$ runs over all vacua) changes over time~\cite{Garriga:2001ri}. The fractional volume in each vacuum is equivalent to the prior $P_q$ defined above. Following the treatment of Linde in Ref.~\cite{Linde:2006nw}, the rate equations generically take the form
\be \dot{f}_i = \dot{Q}_i - \dot{R}_i + \a 3 H^\b f_i.\ee
We should explain the meaning of the various terms: $\dot{Q}_i$ represents the total rate of production of vacuum $i$, while $\dot{R}_i$ represents the total rate at which volume is depleted from vacuum $i$ into other vacua. For a terminal vacuum $j$ (also sometimes called a ``sink''), by definition the corresponding $\dot{R}_j=0.$ The units of volume that the $f_i$ measure, or equivalently, the units of time that the dot in the above expressions stands for, depend on the values of the parameters $\a$ and $\b.$ The most common choices are
\begin{itemize}
\item $\a=0$: this corresponds to using comoving coordinates, and $f_i$ measures the fraction of comoving volume residing in vacuum $i$. Because the total comoving volume of a given region of space is conserved, it is clear that the rate equations must satisfy the consistency condition $\sum_i (\dot{Q}_i - \dot{R}_i) =0.$
\item $\a=1,\, \b=0$: this corresponds to measuring time in units of $H^{-1},$ whatever the Hubble rate in vacuum $i$ happens to be. Thus, the unit of time is in general different in each vacuum. Equivalently, with this choice of parameters time is proportional to $\ln a,$ where $a$ denotes the local scale factor. For this reason, this choice is typically referred to as ``scale-factor time'' (and sometimes also as ``pseudo-comoving''). The factor of $3$ in the last term in the rate equations is simply due to the fact that we are considering three spatial dimensions.
\item $\a=1,\, \b=1$: with this choice of parameters, time is measured in units of proper time, \ie time flows as it would on an ordinary watch. Correspondingly, the volume measured by $f_i$ is then ordinary physical volume.
\end{itemize}

In all cases, the expectation is that in the far future, the multiverse comes to realize the perfect cosmological principle on ``super-bubble'' scales, in that the distribution of vacua becomes homogeneous, isotropic and time-independent. All measures therefore evaluate probabilities in the asymptotic future, where the rate equations are solved by a stationary solution up to exponentially decaying terms (transients). We can evaluate the volume fraction in each vacuum at finite time, and then take the time at which we evaluate the ratio to infinity to obtain the prior probability distribution.

It is also possible to use the rate equations to compute the prior probability distribution for local measures. Here, the quantity of interest is $Q_i= \int \dot{Q}_i dt$ rather than $f_i,$ and the relative probabilities for vacuum $i$ rather than $j$ are defined by $Q_i/Q_j.$. The most prominent example of such a measure is the ``causal diamond'' measure~\cite{Bousso:2006ev}, which uses comoving volumes ($\a=0$). In this case, the ratio of charges is equal  to the relative number of vacuum entries. In Appendix~\ref{sec:Boussorates} we present an alternative calculation of the prior using the methods of Ref.~\cite{Bousso:2006ev}.

\subsection{Basic setup - rate equations with comoving volume weighting} \label{Section-RateEqs-comoving}

By making a few assumptions, we can derive rate equations to tell us about the fraction of volume at different times in different phases during the eternal inflation/cyclic universe hybrid. We define three phases: the first is the false vacuum phase, which we denote by $F.$ The second phase is denoted by $D,$ and it comprises both the dark energy and the ekpyrotic phases. Both of these phases have the effect of smoothing out the universe, and for our purposes it makes sense to group them together. The third phase that we consider is the $R$ phase which describes the radiation and matter dominated epochs during which most stars, galaxies and ordinary observers are created. The $D$ and $R$ phases last for a proper time $t_{D}$ and $t_{R}$ respectively, and the three phases have average Hubble rates $H_{F,D,R}.$ The decay rate $\Gamma_{DF}$ from the false vacuum to the dark energy phase is determined from the CDL instanton. Further, we assume that, for now, the rate $\Gamma_{FD}$ of up-tunneling from $D$ to $F$ is zero. We start off with a parcel of volume completely in the false vacuum, and would like to determine the fraction of comoving volume (defined by the original domain) in each of the phases as a function of time. Under the assumption that each bubble cuts out a comoving Hubble volume (it's asymptotic size in comoving coordinates) when formed~\footnote{Throughout the paper, we work in the ``square bubble approximation," which ignores the difference in the time foliations inside and outside the bubble. As shown in Ref.~\cite{Bousso:2007nd}, this is a good approximation.}, the depletion of volume from the false-vacuum is described by the equation:
\begin{equation}
\frac{df_F}{dt} = - \Gamma_{DF} f_F (t).
\end{equation}
In this paper we will always assume that initially, all the volume resides in the false vacuum, $f_F(0)=1$. With this initial condition, the solution of the equation above is simply
\begin{equation}
f_F = \exp \left[ -\Gamma_{DF} t \right].
\end{equation}

As explained in section \ref{Section-Setup}, we make a distinction between two classes of cyclic models: those for which the entire bubble interior proceeds from the $D$ phase through the bounce to the $R$ phase, and those for which only a fraction $p$ makes it through each crunch/bang transition. We treat the former case first, and then generalize. First, by conservation of comoving volume we know that the rate of depletion of $F$ should be equal to the total rate of growth in $D$ and $R$:
\begin{equation}
\frac{df_F}{dt} = - \left( \frac{df_{D}}{dt}  + \frac{df_{R}}{dt} \right).
\end{equation}
Initially, all of the volume goes from $F$ to $D$, but after a time $t_{D}$, bubbles will begin to go over to the $R$ phase. The rate at which comoving volume is depleted from $D$ is equal to the rate at which volume was added to $D$ from $F$ at time $t - t_{D}$. Then, after a time $t_{D} + t_{R}$, bubbles will begin re-entering $D$. Comoving volume goes from $R$ back to $D$ at a rate equal to the rate volume was added to $D$ at time $t - t_{D} - t_{R}$. The volume in each bubble will then cycle back and forth  between the $D$ and $R$ phases endlessly. Using this information, we conclude that the volume fraction at all times can be found by solving the following set of rate equations:
\begin{eqnarray}\label{eq:req-exact}
\dot{f}_F &=&  -\Gamma_{DF} f_F, \nonumber \\
\dot{f}_{D} &=& \Gamma_{DF} f_F + \sum_{n=1}^{\infty} \sum_{m=n-1}^{n} (-1)^{n+m} \Gamma_{DF} f_F \left( t - n t_{D} -m t_{R} \right) \Theta \left( t - n t_{D} -m t_{R} \right) \\
\dot{f}_{R} &=& - \sum_{n=1}^{\infty} \sum_{m=n-1}^{n} (-1)^{n+m} \Gamma_{DF} f_F \left( t - n t_{D} -m t_{R} \right) \Theta \left( t - n t_{D} -m t_{R} \right), \nonumber
\end{eqnarray}
where $\Theta$ is the heaviside step function.

\begin{figure}
\begin{center}
\includegraphics[width=10cm]{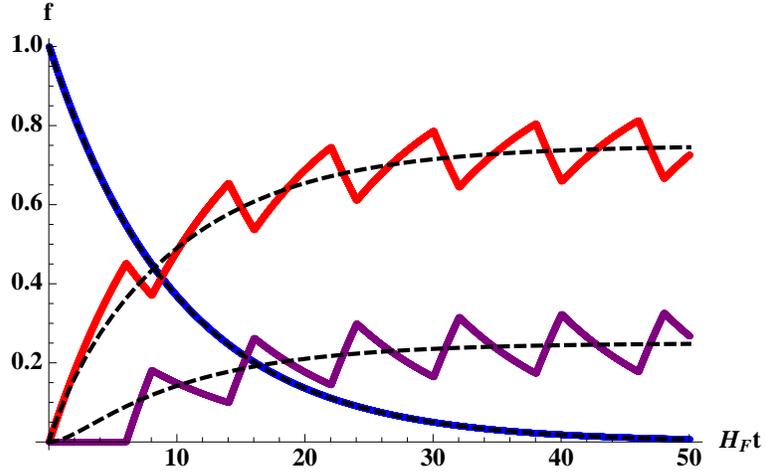}
\caption{A comparison of the solutions to Eq.~\ref{eq:req-exact} (solid) and Eq.~\ref{eq:req-approx2} (dashed) for $t_{D}=6 H_F^{-1}$, $t_{R}=2 H_F^{-1}$, $\Gamma_{DF} =.1 H_F$. The continuum approximation works very well at estimating the average volume in each phase.
\label{fig-no_crunch} }
\end{center}
\end{figure}

In the limit where the lifetime of the false vacuum, $1/\Gamma_{DF}$, is greater than the time spent in either the dark energy/ekpyrotic or radiation/matter phases, we can make an important approximation. Defining the ``decay rates" $\Gamma_{RD} = t_{D}^{-1}$ and $\Gamma_{DR} = t_{R}^{-1}$, we can approximate Eqs.~\ref{eq:req-exact} by the following system of equations:
\begin{eqnarray}\label{eq:req-approx}
\dot{f}_F &=& -\Gamma_{DF} f_F \nonumber \\
\dot{f}_{D} &=& \Gamma_{DF} f_F - \Gamma_{RD} f_{D} + \Gamma_{DR} f_{R} \\
\dot{f}_{R} &=& \Gamma_{RD} f_{D} - \Gamma_{DR} f_{R}. \nonumber
\end{eqnarray}
In Fig.~\ref{fig-no_crunch} we show an example comparing the solutions to the exact and approximate rate equations. It can be seen that the agreement is excellent. In the following, we will mostly use the continuous approximation, where cycles are treated via transition rates. Writing the transition rates out explicitly, we equivalently have
\begin{eqnarray}\label{eq:req-approx2}
\dot{f}_F &=& -\Gamma_{DF} f_F  \nonumber \\
\dot{f}_{D} &=& \Gamma_{DF} f_F - \frac{1}{t_{D}} f_{D} + \frac{1}{t_{R}} f_{R}  \\
\dot{f}_{R} &=& \frac{1}{t_{D}} f_{D} - \frac{1}{t_{R}} f_{R}. \nonumber
\end{eqnarray}

Because $F$ gets depleted continually, at late times $f_F$ tends to zero, $f_{F}(t \rightarrow \infty)=0$.
In the late time limit, it is straightforward to verify that
\be d \equiv \frac{f_{D}(t \rightarrow \infty)}{ f_{R}(t \rightarrow \infty)} = \frac{t_{D}}{t_{R}}.\ee
The volume cycles endlessly between the $D$ and $R$ phases, with the relative proportion in each phase set by the time spent in each.

We will also have occasion to investigate models where only a fraction of the volume makes it through crunch. In this case, assume that the crunch is avoided with probability $p$. Then comoving volume goes from $D$ over to $R$ at a rate equal to the rate volume was added to $D$ at time $t - t_{D}$ multiplied by $p,$ with the rest ending up in a crunch (in other words, all of this volume leaves $D,$ but only a fraction $p$ of it ends up in $R$). To account for the suppression of the volume added to $R$, we should discount the rate at which volume goes back from $R$ to $D$ by a factor of $p$. On the next cycle, we should multiply by $p$, and so on. Therefore, to account for comoving volume lost at each crunch, we should modify Eqs.~\ref{eq:req-exact} as follows:
\begin{eqnarray}\label{eq:req-exact-2}
\dot{f}_F &=&  -\Gamma_{DF} f_F, \\
\dot{f}_{D} &=& \Gamma_{DF} f_F + \sum_{n=1}^{\infty} \sum_{m=n-1}^{n} (-1)^{n+m} p^m \Gamma_{DF} f_F \left( t - n t_{D} -m t_{R} \right) \Theta \left( t - n t_{D} -m t_{R} \right)  \\
\dot{f}_{R} &=& - \sum_{n=1}^{\infty} \sum_{m=n-1}^{n} (-1)^{n+m} p^{n} \Gamma_{DF} f_F \left( t - n t_{D} -m t_{R} \right) \Theta \left( t - n t_{D} -m t_{R} \right).
\end{eqnarray}

\begin{figure}
\begin{center}
\includegraphics[width=10cm]{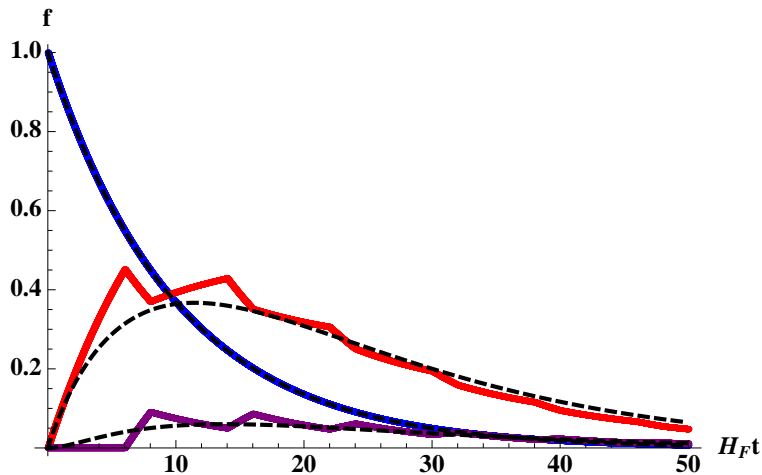}
\caption{A comparison of the solutions to Eq.~\ref{eq:req-exact-2} (solid) and Eq.~\ref{eq:req-approx-2} (dashed) for the same parameters used in Fig.~\ref{fig-no_crunch}, but where only 50\% of the volume makes it through the crunch/bang transition ($p = .5$). We exclude from this plot the volume fraction in the regions that have crunched (which dominates the volume fraction at late times). Again, the continuum approximation works well at determining the volume fractions.
\label{fig-yes_crunch} }
\end{center}
\end{figure}

The continuum approximation is:
\begin{eqnarray}\label{eq:req-approx-2}
\dot{f}_F &=& -\Gamma_{DF} f_F  \\
\dot{f}_{D} &=& \Gamma_{DF} f_F - \frac{1}{t_{D}} f_{D} + \frac{1}{t_{R}} f_{R}\\
\dot{f}_{R} &=& \frac{p}{t_{D}} f_{D} - \frac{1}{t_{R}} f_{R}  \\
\dot{f}_{C} &=& \frac{1-p}{t_{D}} f_{D}.
\end{eqnarray}
Here, we have introduced $f_{C}$, the fraction of comoving volume that ends in a crunch. The continuous approximation is compared to the solution of the discrete equations in Fig.~\ref{fig-yes_crunch}, where it can be seen that the agreement is good. In this example, all of the comoving volume is eaten up by the crunch at late times, regardless of how large $p$ is.

\subsection{Physical volume weighting and the youngness paradox} \label{Section-PhysicalVolWeighting}

The previous section allowed us to set up notation and introduce the rate equations. We can now apply these to the first measure of real interest, namely the proper time cut-off/physical volume weighting measure. In some sense, this is the most natural measure to consider, as (all else being equal) it simply counts probabilities in proportion to the actual physical volume of space that a given phase of interest occupies. Given its simplicity, it is surprising that this measure yields nonsensical results when applied to inflationary bubble universes. This is often referred to in the literature as the ``youngness problem''~\cite{Guth:2000ka}. We will briefly explain this paradox, and then show how the cyclic universe can avoid it.

The false vacuum expands with Hubble rate $H_F.$ Given that nucleation rates are typically very small, this implies that the volume of the false vacuum grows approximately as $e^{3H_F t}.$ When an inflationary bubble is nucleated, it will initially expand at a fast rate during its post-nucleation inflationary phase. However, if this bubble universe is to look like ours, then the vacuum energy must decrease drastically at the end of inflation and relax to the value of the currently measured dark energy, some $100$ orders of magnitude below the inflationary energy density. Forever after the inflationary bubble then expands at the rate of dark energy, \ie it doubles in size every 10 billion years or so. Since the energy density of the false vacuum must be at least as large as the energy density during the inflationary phase (which is typically around the grand unified scale), this implies that the false vacuum expands at an enormously higher rate, doubling every $10^{-30}$ seconds or less!

This implies that every second it becomes $10^{30}$ times more likely for a bubble to nucleate, and the vast majority of bubbles in existence are extremely young. Why does this lead to problematic predictions? Well, you could imagine the entire chain of events that had to take place for you to be able to read this paper: a bubble nucleation, an inflationary phase, early density perturbations evolving into galaxies, the birth of the solar system, the first life-forms on Earth, all of Darwinian evolution, the whole story. Could this whole chain of events have proceeded slightly quicker? Could it have taken just one second less? That certainly seems conceivable, in fact one can guess that the probability for this whole chain of events taking one second less would be roughly equal to the probability of it taking as long as it took. However, the physical volume weighting measure would predict that it would be roughly $10^{30}$ times more likely to have taken one second less, $10^{31}$ times more likely to have taken 10 seconds less,...  The origin of this conclusion is easy to pinpoint: it is a direct consequence of the fact that the false vacuum grows so much faster than the regions harboring ordinary observers like us. As we will now show, for cyclic bubbles this situation can be reversed, and consequently the youngness paradox can be avoided.

In order to weight by physical volume, we simply add a factor $3 H_{F,D,R}$ to the rate equations, where $H_{F,D,R}$ are the average Hubble rates in $F$, $D$ or $R$. The regions that end up in a crunch don't grow, and correspondingly we set their Hubble rates to zero. In the continuous approximation, this yields\footnote{For the discrete rate equations, it would not be enough to add factors of $3 H_{F,D,R}$ to the rate equations, one also has to take into account the growth that occurred between the inception of the bubble and the time $t$ by multiplying the double sum terms in Eq.~\ref{eq:req-exact} by factors of $Exp[3(nH_{D}t_{D}+mH_{R}t_{R})]$.}:
\begin{eqnarray}\label{eq:req-approx2-vol}
\dot{f}_F &=& -\Gamma_{DF} f_F +3H_F f_F \label{eq:req-approx2-vol-F}\\
\dot{f}_{D} &=& \Gamma_{DF} f_F - \frac{1}{t_{D}} f_{D} + \frac{1}{t_{R}} f_{R} + 3 H_{D} f_{D} \label{eq:req-approx2-vol-DE}\\
\dot{f}_{R} &=& \frac{p}{t_{D}} f_{D} - \frac{1}{t_{R}} f_{R} + 3 H_{R} f_{R} \label{eq:req-approx2-vol-RM}\\
\dot{f}_{C} &=& \frac{1-p}{t_{D}} f_{D}. \label{eq:req-approx2-vol-C}
\end{eqnarray}
Conceptually, it is useful to treat the cyclic bubble regions $D,$ $R$ and $C$ together, by defining $f_{CYC} \equiv f_{D} + f_{R} + f_C.$ Then the last three rate equations above can be added to yield \be \dot{f}_{CYC} = \Gamma_{DF} f_F + 3 H_{CYC} f_{CYC}, \label{eq:req-approx2-vol-CYC}\ee and, defining
\begin{equation}
c \equiv \frac{f_{C}(t \rightarrow \infty) }{f_{R}(t \rightarrow \infty)}
\end{equation}
at late times we have
\be
H_{CYC} = \frac{d H_{D} + H_{R}}{1+d+c}.
\ee
With the initial conditions $f_F(t=0)=1, f_{CYC}(t=0) = 0,$ we obtain the solutions
\bea
f_F &=& e^{(3H_F - \Gamma_{DF})t} \label{sol:req-approx2-vol-F} \\
f_{CYC} &=& \frac{\Gamma_{DF}}{3H_F - \Gamma_{DF} - 3H_{CYC}}[e^{(3H_F - \Gamma_{DF})t} - e^{3H_{CYC}t}].\label{sol:req-approx2-vol-CYC}
\eea
In addition, requiring (\ref{eq:req-approx2-vol-RM}) and (\ref{eq:req-approx2-vol-CYC}) to be consistent with each other implies the relation
\be
\frac{pd}{t_{D}} -\frac{1}{t_{R}} + 3 H_{R}= 3\frac{dH_{D} + H_{R}}{1+d+c} + \frac{\Gamma_{DF}}{(1+d+c)}\frac{f_{F}(t \rightarrow \infty)}{f_{R}(t \rightarrow \infty)}. \label{eq:req-consistency}
\ee

There are two cases to consider:

\noindent {\it Case I:} Large false vacuum Hubble rate, $3H_F - \Gamma_{DF} > 3H_{CYC}.$

In that case the solutions above imply that
\be
\frac{f_{F}(t \rightarrow \infty)}{f_{CYC}(t \rightarrow \infty)} = \frac{3H_F - \Gamma_{DF} - 3H_{CYC}}{\Gamma_{DF}}.
\ee
Plugging back into (\ref{eq:req-consistency}) gives
\bea
d = \frac{f_{D}(t \rightarrow \infty)}{ f_{R}(t \rightarrow \infty)} &=& \frac{t_{D}}{p}(3H_F - \Gamma_{DF} - 3H_{R} + \frac{1}{t_{R}}) \\
&\approx& \frac{3t_{D}H_F}{p}.
\eea
Compared to the comoving volume case, the $D$ regions are much larger now compared to the $R$ regions. This is because of the fast expansion of the background, which feeds directly into $D.$ Correspondingly, the crunch regions are small as they do not expand. We can obtain an explicit expression for $c$ by combining the consistency conditions above with Eq. (\ref{eq:req-approx2-vol-C}), which yields
\bea
c  =  \frac{f_{C}(t \rightarrow \infty) }{f_{R}(t \rightarrow \infty)} &=& \frac{(1-p)d}{t_{D}(3H_F-\Gamma_{DF})} \\
&\approx& \frac{1-p}{p}.
\eea
Thus, when $p$ is smaller than $1/2,$ the collapsed regions are actually larger than the radiation/matter regions, which are then completely subdominant. Of course, because of the fast background expansion, which by assumption is faster than the expansion inside the bubbles, there is a youngness paradox in the same way as one would have for inflationary bubbles.

\noindent {\it Case II:} Small false vacuum Hubble rate, $3H_F - \Gamma_{DF} < 3H_{CYC}.$

This is the case of real interest. The solutions (\ref{sol:req-approx2-vol-F})-(\ref{sol:req-approx2-vol-CYC}) imply that for a small false vacuum Hubble rate
\be
\frac{f_{F}(t \rightarrow \infty)}{f_{CYC}(t \rightarrow \infty)} = 0.
\ee
Requiring (\ref{eq:req-approx2-vol-DE}) and (\ref{eq:req-approx2-vol-RM}) to be consistent with each other at late times gives a quadratic equation for $d:$
\be
d^2 + d[\frac{1}{p}-\frac{t_{D}}{pt_{R}}+\frac{3t_{D}}{p}(H_{R}-H_{D})]-\frac{t_{D}}{pt_{R}}=0.
\ee
This equation has one positive and one negative solution, and picking the positive one we obtain
\bea
d &=& -\frac{1}{2}[\frac{1}{p}-\frac{t_{D}}{pt_{R}}+\frac{3t_{D}}{p}(H_{R}-H_{D})] +\frac{1}{2}\sqrt{[\frac{1}{p}-\frac{t_{D}}{pt_{R}}+\frac{3t_{D}}{p}(H_{R}-H_{D})]^2+4\frac{t_{D}}{pt_{R}}}.
\eea
The regime of physical interest is where the Hubble rate in the radiation/matter phase is significantly larger than the other rates in the problem, \ie $H_{R} \gg H_{D},1/t_{R},1/t_{D}.$ Then
\be
d = \frac{f_{D}(t \rightarrow \infty)}{ f_{R}(t \rightarrow \infty)} \approx \frac{1}{3t_{R}H_{R}} \ll 1.
\ee
In other words, the late time ratio of radiation/matter regions to the dark energy/ekpyrosis regions is large $f_{R}(t \rightarrow \infty)/f_{D}(t \rightarrow \infty) \approx 3 t_{R} H_{R}$ and independent of $p.$ For a realistic cyclic universe, the radiation/matter phase lasts about 10 billion years, during which time the universe expands by a factor $e^{60},$ which would give $f_{R}(t \rightarrow \infty)/f_{D}(t \rightarrow \infty) \approx e^{60},$ implying that the radiation/matter phase completely dominates the volume fraction. The false vacuum volume fraction is completely negligible, and so are the collapsed regions: combining (\ref{eq:req-approx2-vol-RM}) and (\ref{eq:req-approx2-vol-C}) gives
\be
c = \frac{f_{C}(t \rightarrow \infty) }{f_{R}(t \rightarrow \infty)} \approx \frac{1-p}{9t_{R}t_{D}H_{R}^2} \lll 1.
\ee
These results hold as long as the bubbles grow faster than the surrounding false vacuum $3 H_{CYC} > 3H_F - \Gamma_{DF},$ or $H_{R} \gtrapprox H_F.$ Thus, using numerical values appropriate to our own universe, these results hold as long as the false vacuum energy density stays below $(10 GeV)^4$ or so.

The most important implication is that the cyclic universe, even with the modification that the dark energy plateau contains a false vacuum (as envisioned in Fig. \ref{Fig-CyclicPlusFVEI}), does not suffer from the youngness paradox. On the contrary, the most likely place to be is in the radiation/matter phase in a late cycle inside an old cyclic bubble.

\subsection{Causal diamond measure and Boltzmann Brains} \label{Section-CausalDiamondBB}

Having just discussed the prototypical global measure, namely weighting by physical volume, we now turn our attention to the prototypical local measure, namely the causal diamond measure~\cite{Bousso:2006ev}. As briefly explained at the beginning of this section, the idea behind the causal diamond measure is to follow a single worldline, and count the number of times different vacua are entered. A closely related prescription would be to count the transitions themselves~\cite{Aguirre:2006na}. One motivation for counting only vacuum entries (as emphasized by Linde~\cite{Linde:1993xx,Linde:2006nw}) is that for inflationary bubbles, only the spacetime region immediately following the phase of inflation can harbor ordinary observers, the entire future part of the bubble is simply empty de Sitter space. And it does not particularly matter just how big this empty de Sitter region is, if one cannot live there anyway.

The causal diamond measure suffers from the ``Boltzmann brain'' problem. As just described, inside inflationary bubbles ordinary observers are formed shortly after the end of inflation. Then, dark energy comes to dominate, and the age of normal observers ends. The only way to re-create normal observers is for a Hubble patch to tunnel up to the false vacuum, and then nucleate a new inflationary bubble. This process is extraordinarily rare. It is exponentially  more likely that a smaller fluctuation occurs, producing a galaxy, solar system, planet, or even a single disembodied brain; anything will be more probable than going all the way back to the false vacuum. Thus, such ``freak'' observers, called Boltzmann brains (BBs), will proliferate without bound. This is the BB problem.

Since BBs arise as thermal fluctuations, they can only exist in regions where the cosmological constant is positive. In Minkowski space for example (which typically arises as a supersymmetry-preserving solution in string theory), such freak observers do not exist. Thus, if all regions of positive cosmological constant decay into zero- or negative-cosmological constant spacetimes on a timescale that is shorter than that of producing a BB, the BB problem does not arise~\cite{Page:2006dt}. Whether all possible positive energy vacua (in particular within string theory) satisfy this requirement is unknown at present (though see ~\cite{Freivogel:2008wm}).

Inside a cyclic bubble formed from an eternally inflating parent vacuum, there will be no chance for Boltzmann brains to form. The timescale for producing a BB is far longer than the duration of the dark energy dominated phase. However, since we are contemplating fairly low energy scales for the false vacuum (our only requirement is that the false vacuum Hubble rate is larger than it is currently), we might worry about freak observers that form there.

To stick to the spirit of the causal diamond measure, when considering the cyclic universe, we should count each entry into the radiation/matter phase separately, as if it were a different vacuum. In addition, each encounter with a BB should be counted as the entry into a new vacuum. Thus we introduce two new vacua $B1$ and $B2$, each corresponding to the false vacuum with a BB. These vacua are cycled between on a timescale $t_B$. In order to model the decay to Minkowski (or AdS) space, we allow all three (physically equivalent) vacua $F,B1,B2$ to decay to a terminal vacuum $S$ with a rate $\Gamma_{SF}.$ The interrelationship of these vacua is shown in Fig. \ref{Fig-CyclicVacua}, and the corresponding rate equations are:

\begin{figure}
\begin{center}
\includegraphics[width=0.75\textwidth]{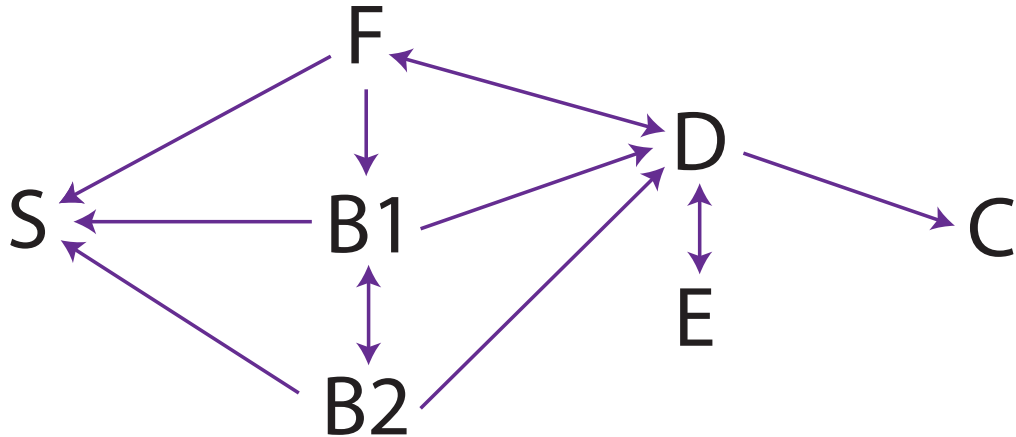}
\caption{Relationships between the different vacua that we consider in this section.}
\label{Fig-CyclicVacua}
\end{center}
\end{figure}

\bea
\dot{f}_F &=& -\Gamma_{DF}f_F + \Gamma_{FD} f_{D} - \frac{1}{t_B}f_F - \Gamma_{SF} f_F \label{eq:req-BB-F} \\
\dot{f}_{B1} &=& \frac{1}{t_B} f_F + \frac{1}{t_B}f_{B2} - \frac{1}{t_B} f_{B1} - \Gamma_{SF} f_{B1}- \Gamma_{DF}f_{B1}  \label{eq:req-BB-B1} \\
\dot{f}_{B2} &=& \frac{1}{t_B} f_{B1} - \frac{1}{t_B} f_{B2} -\Gamma_{SF} f_{B2} - \Gamma_{DF}f_{B2} \label{eq:req-BB-B2} \\
\dot{f}_S &=& \Gamma_{SF} (f_F + f_{B1} + f_{B2}) \label{eq:req-BB-S} \\
\dot{f}_{D} &=& \Gamma_{DF} (f_F +f_{B1} + f_{B2})- \Gamma_{FD} f_{D} -\frac{1}{t_{D}} f_{D} + \frac{1}{t_{R}} f_{R} \label{eq:req-BB-DE} \\
\dot{f}_{R} &=& \frac{p}{t_{D}} f_{D} - \frac{1}{t_{R}} f_{R} \label{eq:req-BB-RM} \\
\dot{f}_C &=& \frac{1-p}{t_{D}} f_{D}. \label{eq:req-BB-C} 
\eea
We are only interested in counting entries into various vacua, hence we are only interested in those terms in the rate equations which reside at the right hand side of the equations and enter with a plus sign. Since we would like to compare ordinary observers (residing in $R$) to BBs (residing in B1 and B2), the ``incoming probability currents'' of interest are
\bea
\dot{Q}_{R} &=& \frac{p}{t_{D}}f_{D} \\
\dot{Q}_{B1} &=& \frac{1}{t_B} (f_F + f_{B2}) \\
\dot{Q}_{B2} &=& \frac{1}{t_B} f_{B1}
\eea
Defining the total ``charge'' $Q_{BB} \equiv Q_{B1}+Q_{B2},$ we are thus interested in the ratio of charges
\be \label{eq:BBratio1}
\frac{Q_{R}}{Q_{BB}} = \frac{p t_B}{t_{D}}\frac{\int^\infty f_{D}}{\int^\infty (f_F + f_{B1} + f_{B2})}.
\ee
Our initial condition is that all of the volume resides in the false vacuum $F$. In the far future, all the comoving volume will be in the terminal vacua $S$ and $C$. Thus, the non-terminal vacua $B1,B2,D,R$ have vanishing comoving volume fraction both initially and at late times, and in integrating the left hand sides of Eqs. (\ref{eq:req-BB-B1}), (\ref{eq:req-BB-B2}), (\ref{eq:req-BB-DE}) and (\ref{eq:req-BB-RM}) from the initial time up to infinity, we find that these integrals are zero. The right hand sides of the integrated Eqs. \ref{eq:req-BB-DE} and \ref{eq:req-BB-RM} lead to
\be
\Gamma_{DF} \int^\infty (f_F + f_{B1} + f_{B2}) = (\Gamma_{FD} + \frac{1-p}{t_{D}}) \int^\infty f_{D}.
\ee
Substituting into Eq.~\ref{eq:BBratio1}, we immediately obtain:
\be
\frac{Q_{R}}{Q_{BB}} = \frac{p t_B \Gamma_{DF}}{(1-p+t_{D}\Gamma_{FD})}.
\ee
For completeness, we show that the same result can be obtained using Bousso's original matrix method~\cite{Bousso:2006ev} -- see appendix B for the derivation. There are a number of things to note about this relation. First, the only effect of the sink $S$ seems to be to justify sending the volume fractions to zero at late times when $p \neq 1$; the result is independent of $\Gamma_{SF}$. The factor $t_{D}\Gamma_{FD}$ in the denominator is expected to be extremely tiny, since it compares a timescale a few orders of magnitude larger than the current age of the universe to a (immensely larger) timescale for up-tunneling from $D$ to $F$. Thus, the BB problem is avoided as long as
\begin{equation}
t_B \Gamma_{DF} > \frac{(1-p+t_{D}\Gamma_{FD})}{p}.
\end{equation}
When $p=1,$ this condition is clearly satisfied~\footnote{This result is in agreement with that of Ref.~\cite{Aguirre:2006ak}, which showed that tightly coupled vacua that cycle between each other quickly dominate this probability measure.}. When $p$ is not exponentially close to one, this condition amounts to requiring that the false vacuum decays to cyclic bubbles faster than it nucleates BBs. This is the standard requirement for a BB problem to not arise in this measure. 

We conclude that theories where $p=1$ will robustly avoid a BB problem, while theories with $p \neq 1$ must have a sufficiently fast decay rate to avoid a BB problem. It is interesting to see that when $p\neq 1$ the rate of production of ordinary observers inside cyclic bubbles becomes irrelevant, as a consequence of the argument given above: in terms of comoving volume, the late-time cycles inside of cyclic bubbles become insignificant. Consequently, if we assume that it is faster to down-tunnel from the false vacuum than to fluctuate a BB, the most likely place to live, according to this measure, is as an ordinary observer in an early cycle inside of a cyclic bubble universe.

\section{Including Inflationary Bubbles}\label{sec:inflationarybubbles}

So far, we have focussed on the cyclic model of the universe, considered by itself. Our main addition, compared to previous discussions of the cyclic universe, was to allow for a phase of false vacuum eternal inflation to coexist with the cyclic universe, and in this context we have shown that the cyclic universe does not necessarily lead to the probability paradoxes that can occur for false vacuum eternal inflation combined with exclusively inflationary bubbles. In the present section we will consider the case where inflationary bubbles and cyclic bubbles coexist.

The motivation for doing so is that, if the string theory landscape exists, and if both inflationary and cyclic solutions are allowed as vacua of string theory, then eternal inflation will physically realize both cosmologies. In that case, one may of course wonder which type of universe we are more likely to inhabit on theoretical grounds~\footnote{Hopefully we will also know soon (perhaps already from results of the PLANCK satellite) which type of universe is favored by observational data.}.

\subsubsection{Physical volume weighting}

When considering physical volume weighting, we ignore the subdivision of the false vacuum into $F,B1,B2$ regions, and just denote the false vacuum as the $F$ region again. Likewise, we can ignore up-tunneling from the inflationary and cyclic bubbles to the false vacuum. Our experience from section \ref{Section-PhysicalVolWeighting} further suggests that we can ignore the crunch regions inside of cyclic bubbles, and that correspondingly we can set $p=1.$ Inside each of the inflationary bubbles, we include two phases: actively inflating volume is denoted by the subscript $I$ (with inflation lasting a duration $t_I$) and volume in the post-inflationary phase denoted by $P$.

The rate equations, for physical volume weighting, then read
\begin{eqnarray}\label{eq:req-infcyc-vol}
\dot{f}_F &=& -(\Gamma_{IF} + \Gamma_{DF}) f_F  + 3 H_F f_F \label{eq:req-infcyc-vol-F}\\
\dot{f}_I &=& \Gamma_{IF} f_F -\frac{1}{t_I} f_I+ 3 H_I f_I \label{eq:req-infcyc-vol-I}\\
\dot{f}_P &=& \frac{1}{t_I} f_I+ 3 H_P f_P              \\
\dot{f}_{D} &=& \Gamma_{DF} f_F - \frac{1}{t_{D}} f_{D} + \frac{1}{t_{R}} f_{R} + 3 H_{D} f_{D} \label{eq:req-infcyc-vol-DE}\\
\dot{f}_{R} &=& \frac{1}{t_{D}} f_D - \frac{1}{t_{R}} f_{R} + 3 H_{R} f_{R}. \label{eq:req-infcyc-vol-RM}
\end{eqnarray}
We can lump together $P$ and $I$ into $f_{INF} = f_I + f_P,$ and the $D$ and $R$ regions into cyclic bubbles with volume fraction $f_{CYC} = f_{DE} + f_{RM}$ and corresponding rate equations
\bea \dot{f}_{INF} &=& \Gamma_{IF} f_F + 3 H_{INF} f_{INF} \\ \dot{f}_{CYC} &=& \Gamma_{DF} f_F + 3 H_{CYC} f_{CYC},\label{eq:req-infcyc-vol-CYC}\eea with \be H_{INF} = \frac{iH_I + H_P}{1+i} \qquad H_{CYC} = \frac{d H_{D} + H_{R}}{1+d},\ee where we have defined the late-time ratios \be i \equiv \frac{f_I(t \rightarrow \infty)}{f_P(t \rightarrow \infty)} \qquad d \equiv \frac{f_D(t \rightarrow \infty)}{f_R(t \rightarrow \infty)}. \ee With the initial conditions $f_F(t=0)=1, f_{INF}(t=0)=0, f_{CYC}(t=0) = 0,$ we obtain the solutions \bea f_F &=& e^{(3H_F -\Gamma_{IF}- \Gamma_{DF} )t} \label{sol:req-infcyc-vol-F} \\ f_{INF} &=& \frac{\Gamma_{IF}}{3H_F -\Gamma_{IF}- \Gamma_{DF} - 3H_{INF}}[e^{(3H_F -\Gamma_{IF}- \Gamma_{DF})t} - e^{3H_{INF}t}]\label{sol:req-infcyc-vol-I}\\ f_{CYC} &=& \frac{\Gamma_{DF}}{3H_F -\Gamma_{IF}- \Gamma_{DF} - 3H_{CYC}}[e^{(3H_F -\Gamma_{IF}- \Gamma_{DF})t} - e^{3H_{CYC}t}].\label{sol:req-infcyc-vol-CYC} \eea In addition, and analogously to the case where there are only cyclic bubbles, we have the consistency conditions \bea \frac{i}{t_I}+3 H_P &=& + 3 H_{INF}+ \frac{\Gamma_{IF}}{(1+i)}\frac{f_{F}(t\rightarrow \infty)}{f_{P}(t \rightarrow \infty)} \\ \frac{d}{t_{D}} -\frac{1}{t_{R}} + 3 H_{R} &=& 3H_{CYC} + \frac{\Gamma_{DF}}{(1+d)}\frac{f_{F\infty}}{f_{RM\infty}}. \label{eq:req-infcyc-consistency2}\eea
The case of physical interest is where the false vacuum energy density is larger than the energy density in the inflationary phase, so that we may assume $3H_F -\Gamma_{IF}- \Gamma_{DF} > 3H_{INF}.$ In that case \be \frac{f_{F}}{f_{INF}}(t \rightarrow \infty) = \frac{3H_F -\Gamma_{IF}- \Gamma_{DF} - 3H_{INF}}{\Gamma_{IF}}. \ee The assumption that the false vacuum lies above the inflationary scale implies that it will also lie above $H_{R}.$ Thus we have \be \frac{f_{F}}{f_{CYC}}(t \rightarrow \infty) = \frac{3H_F - \Gamma_{IF}- \Gamma_{DF} - 3H_{CYC}}{\Gamma_{DF}}. \ee Plugging into the consistency relations gives \bea i &=& t_I (3H_F -\Gamma_{IF}- \Gamma_{DF} - 3H_{P} ) \approx 3 t_I H_F \\d &=& t_{D}(3H_F -\Gamma_{IF}- \Gamma_{DF} - 3H_{R} + \frac{1}{t_{R}}) \approx 3 t_{D} H_F.\eea With the assumptions above, we have the approximate late time ratios \bea \frac{f_{F}}{f_{P}}(t \rightarrow \infty) &=& (1+i)\frac{f_{F}}{f_{INF}}(t \rightarrow \infty)\approx \frac{9t_IH_F^2}{\Gamma_{IF}}\\  \frac{f_{P}}{f_{R}}(t \rightarrow \infty) &\approx& \frac{t_D}{t_I}\frac{\Gamma_{IF}}{\Gamma_{DF}}  . \eea The false vacuum vastly dominates over everything else. For comparable nucleation rates, the $P$ regions dominate over the $R$ regions by a huge factor $t_D/t_I.$ This is because of the rapid background expansion, which favors first of all the false vacuum, and secondly those vacua that it directly and most rapidly feeds into. Thus, adding inflationary bubbles can re-introduce the youngness problem.

\subsection{Causal diamond measure}

The extension of the causal diamond probability calculations to the case where inflationary bubbles are included is rather straightforward. It is important to include the possibility of up-tunneling, both from the inflationary bubbles and from the cyclic ones. We also include the possibility that the inflationary bubbles can decay into the sink $S$ with rate $\Gamma_{SP}.$ This time, we only consider Boltzmann brains in the inflationary bubbles, as the false vacuum Hubble rate must necessarily be very high (implying that the horizon in the false vacuum is far too small to harbor a BB). Thus, the BBs of interest principally form in the inflationary bubbles, after inflation has ended and dark energy has taken over. To model this, we once again employ the ``epi-cycle'' vacua $B1,B2,$ where now the vacua $P,B1,B2$ are physically identical. The full rate equations now read
\bea
\dot{f}_F &=& -\Gamma_{DF}f_F - \Gamma_{IF} f_F + \Gamma_{FP}(f_P+f_{B1} + f_{B2})+ \Gamma_{FD} f_{D}  \label{eq:req-infcyc-BB-F} \\
\dot{f}_I &=& \Gamma_{IF} f_F -\frac{1}{t_I}f_I \label{eq:req-infcyc-BB-I} \\
\dot{f}_P &=& \frac{1}{t_I}f_I - \frac{1}{t_B}f_P - \Gamma_{SP} f_P - \Gamma_{FP} f_P  \label{eq:req-infcyc-BB-P} \\
\dot{f}_{B1} &=& \frac{1}{t_B} f_P + \frac{1}{t_B}f_{B2} - \frac{1}{t_B} f_{B1} - \Gamma_{SP} f_{B1}- \Gamma_{FP} f_{B1} \label{eq:req-infcyc-BB-B1} \\
\dot{f}_{B2} &=& \frac{1}{t_B} f_{B1} - \frac{1}{t_B} f_{B2} -\Gamma_{SP} f_{B2} - \Gamma_{FP} f_{B2}\label{eq:req-infcyc-BB-B2} \\
\dot{f}_S &=& \Gamma_{SP} (f_P + f_{B1} + f_{B2}) \label{eq:req-infcyc-BB-S} \\
\dot{f}_{D} &=& \Gamma_{DF} f_F - \Gamma_{FD} f_{D} -\frac{1}{t_{D}} f_{D} + \frac{1}{t_{R}} f_{R} \label{eq:req-infcyc-BB-D} \\
\dot{f}_{R} &=& \frac{p}{t_{D}} f_{D} - \frac{1}{t_{R}} f_{R} \label{eq:req-infcyc-BB-R} \\
\dot{f}_C &=& \frac{1-p}{t_{D}} f_{D} \label{eq:-infcycreq-BB-C},
\eea
and the incoming probability currents of interest are
\bea
\dot{Q}_P &=& \frac{1}{t_I} f_I \\
\dot{Q}_{R} &=& \frac{p}{t_{D}}f_{D} \\
\dot{Q}_{BB} = \dot{Q}_{B1}+\dot{Q}_{B2} &=& \frac{1}{t_B} (f_P + f_{B1} + f_{B2}).
\eea
The questions we are interested in are: what is the relative probability of being an observer inside a cyclic bubble rather than inside an inflationary bubble? What is the ratio of ordinary observers to BBs? The answer to the first question can be expressed as
\be
\frac{Q_{R}}{Q_P} = \frac{pt_I}{t_{D}}\frac{\int^\infty f_{D}}{\int^\infty f_I}.
\ee
As in section \ref{Section-CausalDiamondBB}, the integration of Eqs. \ref{eq:req-infcyc-BB-D} and \ref{eq:req-infcyc-BB-R} implies that
\be
\Gamma_{DF} \int^\infty f_F = (\Gamma_{FD} + \frac{1-p}{t_{D}}) \int^\infty f_{D}.
\ee
Combining this with the integral of Eq. (\ref{eq:req-infcyc-BB-I}) we obtain
\be \frac{Q_{R}}{Q_P} = \frac{p\Gamma_{DF}}{\Gamma_{IF}(1-p+t_{D}\Gamma_{FD})}.
\ee
When the entire cyclic universe makes it through the crunch, we obtain
\be \frac{Q_{R}}{Q_P} = \frac{\Gamma_{DF}}{\Gamma_{IF}t_{D}\Gamma_{FD}}, \qquad (p=1).
\ee
We may assume that the down-tunneling rates $\Gamma_{DF}$ and $\Gamma_{IF}$ are comparable, so that the ratio above becomes equivalent to a semi-classical up-tunneling timescale divided by the classical timescale of the cyclic universe, and hence this ratio is enormous. In this case, it is vastly more likely to live in a cyclic universe than in an inflationary one. This result is again intuitively clear: the causal diamond measure rewards the repeated, infinite production of ordinary observers that occurs inside of cyclic bubbles, compared to the one-time production of ordinary observers inside inflationary bubbles.

When the cyclic universe loses comoving volume at each crunch ($p\neq 1$), the situation changes, since now cyclic bubbles lose their advantage. As described above, at late times the depletion of comoving volume inside cyclic bubbles means that they are essentially treated in the same way as inflationary bubbles, and this is reflected by the approximation \be
\frac{Q_{R}}{Q_P} \approx p\frac{\Gamma_{DF}}{\Gamma_{IF}}, \qquad (p\neq1).
\ee
Now the probabilities of being in a cyclic or an inflationary universe have become comparable, and depend on the details of the tunneling potential and the severeness of the instability of the cyclic model under consideration.

We can also compare the likelihood of being an ordinary observer rather than a BB by considering the ratio
\bea
\frac{Q_P+Q_R}{Q_{BB}} &=& \frac{t_B}{\int^\infty(f_P + f_{B1} + f_{B2})}\Big( \frac{1}{t_I}\int^\infty f_I + \frac{p}{t_D}\int^\infty f_D \Big) \nn \\
&=& \frac{\int^\infty f_P}{\int^\infty (f_P + f_{B1} + f_{B2})}\Big( [1+t_B(\Gamma_{SP} + \Gamma_{FP})]+\frac{p\Gamma_{DF}[1+t_B(\Gamma_{SP} + \Gamma_{FP})]}{\Gamma_{IF}(1-p+t_D\Gamma_{FD})} \Big) \nn \\
&=& t_B(\Gamma_{SP} + \Gamma_{FP})\Big( 1+ \frac{Q_{R}}{Q_P} \Big),
\eea
where we have used the integrals of Eqs. (\ref{eq:req-infcyc-BB-I})-(\ref{eq:req-infcyc-BB-B2}). In the absence of cyclic bubbles, we would have obtained a ratio proportional to $t_B \Gamma_{SP}$, and reached the conclusion that to avoid a BB problem the inflationary bubbles must decay faster than BBs are formed. When the observers inside of cyclic bubbles dominate, as they do when $p=1$ or when the rate to form cyclic bubbles is high enough, the BB problem can be solved even for slow decays of the inflationary bubbles to the sink. To illustrate this, assume that $p=1$, in which case we obtain:
\begin{equation}
\frac{Q_P+Q_R}{Q_{BB}} \simeq \frac{t_B}{t_D} \frac{\Gamma_{DF}}{\Gamma_{FD}} \frac{\Gamma_{SP}}{\Gamma_{IF}}
\end{equation}
As discussed above, the first two ratios are huge, implying that there will be no BB problem as long as the decay to the sink is not extremely slow.

\section{Discussion}\label{sec:conclusions}

In the standard picture of eternal inflation, different phases are seeded from a rapidly inflating parent vacuum through the formation of bubbles, each of which can contain an inflationary cosmology, and possibly describe our observable universe. As we have described in detail above, there exist many ambiguities and subtleties in defining a probability measure suitable for making predictions in an eternally inflating universe. However, many of the problems that arise for various measures can be traced back to a simple fact: during eternal inflation it is much easier to fluctuate to a lower energy state than a higher energy state. The youngness problem arises because space is inflating much more rapidly outside of pocket universes than inside. The Boltzmann Brain problem arises because it is much easier to make observers via a small fluctuation than to re-play the entirety of cosmological evolution.

The model we have introduced in this paper, which incorporates cyclic universes into the picture of eternal inflation, at face value would seem to avoid these problems. In a cyclic bubble universe, cosmological evolution leads periodically to states with higher energy density. Thus, in forming a cyclic bubble, a chain of events is set into motion by which the energy inside the bubble can temporarily exceed that of the background space it is embedded in. The volume inside the bubble can then  grow faster than on the outside, relieving the youngness problem. Cyclic bubbles continually produce observers by normal cosmological evolution, and the nucleation rate for cyclic bubbles can be fast,  alleviating the Boltzman Brain problem. A similar conclusion would be reached in models where the production of false vacuum (higher energy density) bubbles is not highly suppressed, as in the ``de Sitter equilibrium cosmology" of Albrecht~\cite{Albrecht:2004ke,Albrecht:2009vr}. These two examples are illustrative of what is generally necessary to avoid a Youngness and BB problem, and suggest that something like a rather drastic violation of the null energy condition is required. Whether or not this is reasonable is of course subject to debate.

The analysis we have performed in this paper confirms the intuition described above. However, the conclusions are somewhat sensitive to the details of the physics connecting the big crunch of one cycle to the big bang of the next. When there are only cyclic and terminal vacua, the volume fraction of the universe is always dominated by regions containing matter and radiation at late times. Thus, the youngness paradox is non-existent. Again, when there are only cyclic and terminal vacua, the existence of a BB problem is dependent on whether there is any loss of volume at the crunch/bang transition. When there is no loss of volume, the BB problem is non-existent. When there is volume loss, the false vacuum must produce cyclic bubbles faster than it produces BBs to avoid a problem. Let us emphasize that there is still a measure problem for the eternal inflation/cyclic universe hybrid; there are many possible cutoff procedures, and no {\it a priori} way to choose which is correct. However, our results suggest that many commonly considered choices lead to predictions consistent with observations.

We have also analyzed a model that contains both cyclic and inflationary bubbles formed from the same false vacuum. This might be expected more generally if both inflationary and cyclic cosmologies are realized in a fundamental theory. We found that when weighting vacua by physical volume the youngness paradox is re-introduced if the false vacuum energy scale is large and inflationary bubbles are produced more rapidly than cyclic bubbles. This is the regime where the volume fraction inside inflationary bubbles is larger than the volume in the radiation/matter dominated portions of cyclic bubbles. In the causal diamond measure, the cyclic bubbles generically dominate the measure as long as they are produced more frequently than the inflationary bubbles. In the case where there is no loss of volume at the crunch/bang transition, cyclic bubbles in fact always dominate. In cases where the cyclic bubbles dominate the measure, the BB problem is largely absent, although subject to some conditions on the rate of decay of the false vacuum.

Beyond the particular examples we have considered, our results motivate the construction of a more general set of conditions necessary for particular measures to yield results consistent with our observable universe. We hope to explore this question in future work.

\acknowledgements

We would like to thank Adam Brown for his collaboration at the early stages of this work. We have greatly benefitted from discussions with Andy Albrecht, Jens Niemeyer and Paul Steinhardt. J.L.L. would like to express his thanks to Perimeter Institute for hospitality while this work was initiated. J.L.L. gratefully acknowledges the support of the European Research Council in the form of a Starting Grant. Research at Perimeter Institute is supported by the Government of Canada through Industry Canada and by the Province of Ontario through the Ministry of Research and  Innovation.

\appendix

\section{Two-Field Tunneling and Quantum Fluctuations}\label{sec:twofieldbubbles}

\begin{figure}
\begin{center}
\includegraphics[width=10cm]{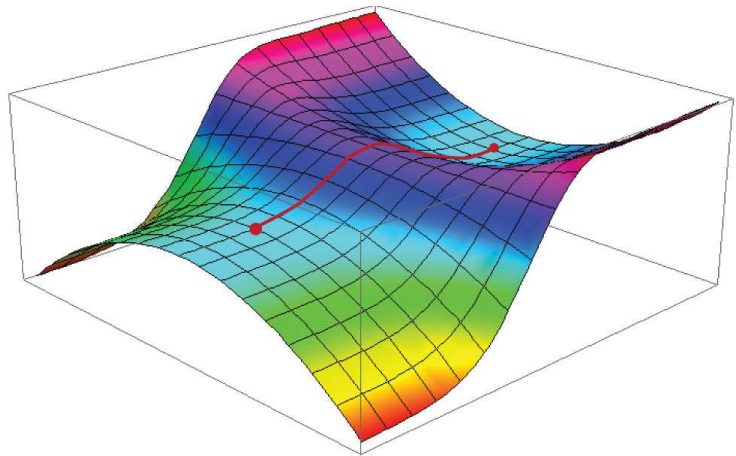}
\caption{\label{figuretunnelingpotential} {\small The tunneling path from a false vacuum (on the right) to the two-field ekpyrotic phase (on the left).
}}
\end{center}
\end{figure}

Fig. \ref{figuretunnelingpotential} illustrates the two-field potential in the tunneling region. The two scalars are $\s$ and $s$, with $\s$ being oriented along the ekpyrotic ridge, and $s$ the coordinate transverse to $\s.$ Up to a shift in $\s,$ and to quadratic order in $s$, the potential after tunneling is \be V = V_0 - V_0 e^{\sqrt{2\e}\s} (1+\e s^2),\ee where in realistic models $\e \sim {\cal O}(10^2)$ (``realistic'' meaning that the spectral tilt then turns out to be at the percent level).

\begin{figure}
\begin{center}
\includegraphics[width=10cm]{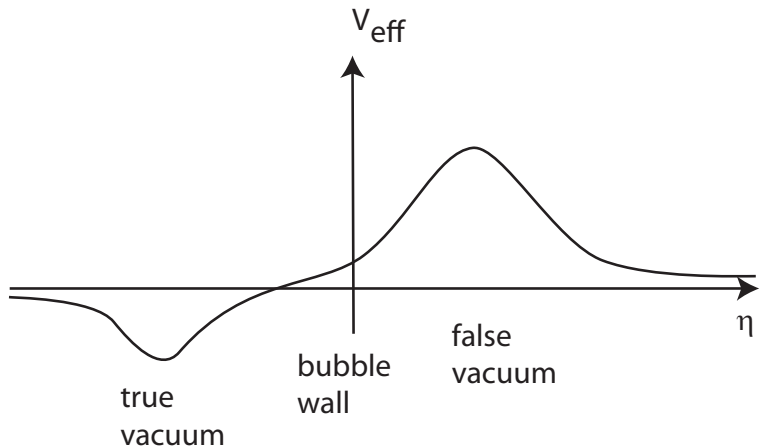}
\caption{\label{figureeffectivepotential} {\small A sketch of the effective potential defined in Eq. (\ref{eq:effectivepotential}).
}}
\end{center}
\end{figure}

The dominant channel for vacuum decay corresponds to the instanton solution with smallest action. This means that the preferred tunneling path is to the ridge ($s=0$), and $\dot\s = 0$ right after bubble nucleation. Since in the vicinity of the ridge we have that $V_{,\s} \gg V_{,s},$ the classical trajectory after tunneling corresponds to slowly rolling down the ridge in the potential. If the entire bubble were to do that, then the subsequent phase of dark energy domination, followed by the 120 e-folds of ekpyrosis, would ensure that the spacetime inside the bubble was sufficiently flat right before the crunch, so that (with the usual assumptions made in the cyclic model) all of it would make it through the crunch/bang transition. Whether or not this is the case depends on the quantum fluctuations of the transverse field $s$. Following Garcia-Bellido {\it et al.}\cite{GarciaBellido:1997te}, we can estimate the fluctuations in $s$ as follows:

The (outside) bubble geometry is described by the line element
\be ds^2 = d\t^2 + a_E^2(\t)[-d\rho^2 + \cosh^2\rho (d\th^2 + \sin^2\th d\phi^2)],\ee and the scalar profile depends on $\t$ alone. The above coordinates in fact cover only the exterior of the light-cone emanating from the origin $\t=0.$ The interior line element is obtained from the above one by performing an analytic continuation
\be t=-i\t \qquad r=\rho + i\frac{\pi}{2} \qquad a(t)=-ia_E(it).\ee
The equation of motion for the (gauge-invariant) fluctuations $\d s$ is \be \Box \d s - V_{,ss} \d s =0, \ee where $V_{,ss}$ is evaluated on the tunneling solution $\s_0(\t),s=0.$ With the ansatz \be \d s = a_E^{-1}(\t)F_p(\t) Y_{plm}(\rho,\th,\phi)\ee and the introduction of a conformal coordinate $\eta,$ defined via $a_E d\eta = d\t,$ the equation of motion separates into a 3-dimensional Klein-Gordon equation \be ^{(3)}\Box Y_{plm} = (p^2 + 1)Y_{plm}\ee and a Schr\"{o}dinger-type equation \be -\frac{d^2 F_p}{d\eta^2}+a_E^2 [V_{,ss} - \frac{R}{6}]F_p = p^2 F_p, \ee where $R$ denotes the 4-dimensional Ricci scalar. Thus $Y_{plm}$ are mode functions on the $2+1$ dimensional de Sitter space spanned by $\rho,\th,\phi$ (there exist explicit expressions for these mode functions in terms of Legendre functions and spherical harmonics). In the Schr\"{o}dinger equation, the effective potential is \be V_{eff} = a_E^2 [V_{,ss} - \frac{R}{6}],\label{eq:effectivepotential}\ee and its shape is sketched in Fig. \ref{figureeffectivepotential}.

Since $a_E \rightarrow 0$ as $\eta \rightarrow \pm \infty,$ the effective potential tends to zero at both ends. Thus there is a continuous spectrum of modes with positive ``energy'' $p^2.$ However, there are also discrete modes (bound states) with $p^2<0.$ Note that in the true vacuum, in the vicinity of the nucleation point, we have that \be \frac{V_{,ss}}{V} \approx -2\e e^{\sqrt{2\e}\s}.\ee The exponential is very small there, hence this implies that \be \frac{|V_{,ss}|}{R} \sim \frac{|V_{,ss}|}{H^2} \ll 1,\ee and in fact we can use the same approximation throughout to obtain an estimate of the wavefunction of the lowest mode. If the $V_{,ss}$ term is dropped entirely in the Schr\"{o}dinger equation, then there is a (non-normalizable) solution with $p^2=-1$ and $F \propto a_E.$ Once $V_{,ss}$ is taken into account, the solution becomes normalizable and the eigenvalue shifts to the value $p^2 = -1+ \gamma$ with
\be \gamma = \frac{H_F^2}{2}\int d\eta a_E^4(V_{,ss}-\frac{R}{6}+2H_F^2).\ee The lowest mode is (now in terms of the bubble interior coordinates) \be \d s^\gamma_{l=0} \approx \frac{H_F}{2\pi \gamma^{1/2}}\frac{\sinh((1-\gamma)^{1/2}r)}{\sinh r}.\ee Near $r=0,$ we have \be \langle (\d s)^2 \rangle^{1/2} \approx \frac{H_F}{2\pi\gamma^{1/2}}\ee and in the thin-wall limit \be \gamma \approx \frac{2V_{,ssF}}{3H_F^2}+\frac{1}{8}H_F^2R_0^4(V_{,ssT}-V_{,ssF}), \ee where $V_{,ssT}$ and $V_{,ssF}$ stand for the effective masses of the $s$ field in the true and false vacua respectively, and $R_0$ is the size of the bubble at nucleation. If we design the tunneling potential such that $\gamma \approx 1,$ then the correlation length $1/\gamma$ of the fluctuations is roughly one horizon size, with a (bubble) quantum fluctuation amplitude similar to the one of the ordinary (cosmological) quantum fluctuations in $s.$ In this case virtually the entire bubble interior makes it through the first crunch, and even the first cycle is perfectly habitable.

\section{Causal Diamond Measure and Matrix Calculations}\label{sec:Boussorates}

Bousso's measure proposal \cite{Bousso:2006ev} is to consider only the causal patch associated to a single worldline, and count the number of times the worldline intersects a given vacuum. Thus neither volume nor lifetime of a given vacuum matter, only transitions between vacua. Inside a cyclic bubble, each cycle is separated from the next by a big bang, and hence, in keeping with the spirit of this proposal, it makes sense to treat subsequent cycles as separate vacua (moreover, it is conceivable that certain physical properties could evolve from cycle to cycle). We consider the same setup as in section \ref{Section-CausalDiamondBB}. Transitions between vacua are encoded in a matrix $\eta.$ Following \cite{Bousso:2006ev}
we denote by $\kappa_{ij}$ the probability per unit proper time for a geodesic worldline in vacuum $j$ to enter vacuum $i.$ The elements of the transition matrix $\eta$ are simply the normalized versions of these transition probabilities, \ie $\eta_{ij} = \kappa_{ij}/\sum_k \kappa_{kj}.$ This normalization ensures that for non-terminal vacua the total probability to leave the vacuum is 1. Hence, each column lists the normalized probabilities to transition from the corresponding vacuum to the vacuum associated with the row in question. For terminal vacua the probability to leave the vacuum is zero by definition, and thus columns of $\eta$ corresponding to terminal vacua are zero. In our case, the sequence of vacua in rows and columns is $F,B1,B2,S,D,R,C.$ The transition matrix $\eta$ is then
\[ \left( \begin{array}{ccccccc}
0 & 0 & 0 & 0 & \frac{\Gamma_{FD}}{\Gamma_{FD} + 1/t_{D}} & 0 & 0 \\ \frac{1}{1+t_B(\Gamma_{DF}+\Gamma_{SF})} & 0 & \frac{1}{1+t_B(\Gamma_{DF}+\Gamma_{SF})} & 0 & 0 & 0 & 0 \\ 0 & \frac{1}{1+t_B(\Gamma_{DF}+\Gamma_{SF})} & 0 & 0 & 0 & 0 & 0 \\ \frac{\Gamma_{SF}}{(1/t_B + \Gamma_{DF}+\Gamma_{SF})} & \frac{\Gamma_{SF}}{(1/t_B + \Gamma_{DF}+\Gamma_{SF})} & \frac{\Gamma_{SF}}{(1/t_B + \Gamma_{DF}+\Gamma_{SF})} & 0 & 0 & 0 & 0 \\
\frac{\Gamma_{DF}}{(1/t_B + \Gamma_{DF}+\Gamma_{SF})} & \frac{\Gamma_{DF}}{(1/t_B + \Gamma_{DF}+\Gamma_{SF})} & \frac{\Gamma_{DF}}{(1/t_B + \Gamma_{DF}+\Gamma_{SF})} & 0 & 0 & 1 & 0 \\
0 & 0 & 0 &0 &\frac{p}{1+t_{D}\Gamma_{FD}} &0 & 0 \\
0 & 0 & 0 & 0& \frac{1-p}{1+t_{D}\Gamma_{FD}} & 0 & 0\\
\end{array} \right)\]

Starting with an initial probability distribution $P_0 = (1,0,0,0,0,0,0),$ after one iteration this gives $\eta P_0,$ then to $\eta^2 P_0,$ and so on. The late-time un-normalized probability to be in a given vacuum is the sum of all these probabilities, namely \be P = \frac{\eta}{1-\eta}P_0.\ee The inverse of $(1-\eta)$ may be singular (this is the case when there are no terminal vacua/sinks), in which case we can calculate the inverse of $(1-\e\eta),$ and after normalizing probabilities at the end, set $\e=1.$ In our case this complication does not arise, and we immediately obtain that the late-time probability distribution is proportional to
\bea P &\propto& \Big(\Gamma_{DF} \Gamma_{FD} t_{D}, \nn \\ &&(1-p+\Gamma_{FD}t_{D})\frac{1+t_B (\Gamma_{SF}+\Gamma_{DF})}{t_B[2+t_B (\Gamma_{SF}+\Gamma_{DF})]},\nn \\ &&(1-p+\Gamma_{FD}t_{D})\frac{1}{t_B [2+t_B (\Gamma_{SF}+\Gamma_{DF})]},\nn \\ && (1-p+\Gamma_{FD}t_{D}) \Gamma_{SF}   ,\nn \\ && \Gamma_{DF}(1+t_{D}\Gamma_{FD})  ,\nn \\ && p \Gamma_{DF}, \nn \\ && (1-p)\Gamma_{DF}\Big).\eea
Thus, in particular the ratio of entries into the ordinary-observer-containing $R$ vacuum to the Boltzmann brains $B1+B2$ is \be \frac{Q_{R}}{Q_{BB}} = \frac{p t_B \Gamma_{DF}}{1-p+t_{D}\Gamma_{FD}},\ee in agreement with the result derived in section \ref{Section-CausalDiamondBB} using the rate equations.

\bibliography{cyclic_measures}

\end{document}